**Full-Field Damage Monitoring in Architected Lattices Using *In situ* Electrical Impedance Tomography**

*Akash Deep, Andrea Samore, Alistair McEwan, Andrew McBride, and Shanmugam Kumar\**

A. Deep, S. Kumar

Sustainable Multifunctional Materials and Additive Manufacturing (SM$^2$AM) Laboratory, James Watt School of Engineering, University of Glasgow, Glasgow G12 8QQ, UK
E-mail: Msv.Kumar@glasgow.ac.uk

A. Samore, A. McEwan

School of Biomedical Engineering, University of Sydney, Sydney, New South Wales, Australia

A. McBride

Glasgow Computational Engineering Centre, James Watt School of Engineering, University of Glasgow, Glasgow G12 8QQ, UK

**ABSTRACT**

Electrical impedance tomography (EIT) enables non-invasive, spatially continuous reconstruction of internal conductivity distributions, providing full-field sensing beyond conventional point measurements. Here, we report the first *in situ* implementation of EIT within a tunable architected lattice materials framework, enabling systematic exploration across a broad lattice design space while achieving real-time monitoring of damage evolution, including early-stage, pre-fracture events, in 3D-printed multifunctional lattice composites. Lattices are designed via Voronoi-based branch–trunk–branch motifs inspired by 2D wallpaper symmetries and fabricated using CNT-infused photocurable resins, with nanoscale filler dispersion confirmed by field-emission scanning electron microscopy. Sixteen electrodes distributed along the lattice periphery enable EIT measurements during quasi-static tensile loading. Conductivity maps reconstructed using adjacent and across current injection schemes resolve sequential ligament fracture with high temporal resolution, with localised conductivity loss quantitatively coinciding with fracture sites, including regions remote from electrodes. Architectural tunability allows systematic control of EIT imaging sensitivity to early-stage damage, while pronounced resistance discontinuities at failure further corroborate spatial localisation; global end-to-end resistance measurements complement macroscopic stress–strain responses. Collectively, these results establish *in situ* EIT as a scalable, full-field sensing modality for architected multifunctional materials, providing an experimentally validated pathway toward autonomous, intelligent materials and data-rich material states that can inform digital twin frameworks for structural, biomedical, and energy-related applications.

**1. Introduction**

Recent advances in additive manufacturing (AM) enable precise fabrication of complex geometries across multiple length scales, opening new opportunities for architected and metamaterials design [1–4]. Metamaterials derive their properties primarily from architecture rather than composition, and when engineered across multiple physical domains, they can achieve multifunctionality, responding to mechanical, electrical, thermal, and optical stimuli [5,6]. These materials enable applications in sensing [7], actuation [8], vibration and sound absorption [9,10], thermal management and insulation [11], medical technologies [12–14], and even cloaking[15]. AM facilitates direct realisation of intricate architectures, including ordered and disordered cellular structures, with high precision, tunability, and reproducibility. The synergy of rational design and AM fabrication has redefined the development and deployment of advanced materials.

Architected lattices exploit micro- and meso-scale geometry to achieve tailored combinations of stiffness, strength, toughness, and functional response [1]. Embedding functionality—such as self-sensing, or electromagnetic shielding—extends their utility beyond conventional load-bearing roles [2,16,17]. However, as geometric and functional complexity increases, robust real-time characterisation techniques for monitoring evolving internal states remain scarce. Conventional pointwise sensing strategies, while effective in simple structures, cannot provide full-field insight into mechanical or electrical phenomena in complex lattices. Despite extensive exploration of their mechanical and multifunctional performance, the potential of architected lattices to actively enhance full-field sensing through geometric design remains largely untapped.

Piezoresistive sensing is widely used for embedding self-sensing functionality in multifunctional composites [18–21]. It monitors strain-induced changes in electrical resistance, with fractional resistance change ($\Delta R/R$) related to applied strain ($\varepsilon$) via the gauge factor (GF). Conductive nanocomposites—particularly CNT networks, graphene, or doped semiconductors—translate deformation into measurable electrical signals [22–24]. In 3D-printed lattices, piezoresistive behaviour can be tuned via unit cell geometry, filler type, concentration, or spatial distribution [25,26]. While simple, sensitive, and compatible, resistance-based sensing is inherently limited to discrete electrode points, constraining its ability to capture distant damage or track the spatial evolution of failure in real time. Recent multiphysics models combined with experiments have captured evolving resistance fields in 2D lattice composites [26] though the approach is complex, and additive-manufacturing-induced errors further limit predictive accuracy [26,27].

Electrical impedance tomography (EIT) provides a full-field, non-invasive alternative for monitoring internal conductivity [28]. By applying electrical currents and measuring boundary voltage differences, EIT reconstructs internal conductivity maps via computational inversion [29]. Unlike piezoresistive sensors, which report localised resistance changes, EIT enables spatially continuous imaging across entire structures, capturing both nearby and distant damage in real time [30–36]. Originally developed for biomedical diagnostics [28,37], EIT has also been applied to process monitoring [38–40], geophysics [41], electronic skins [42–44], non-destructive evaluation [45–47], and structural health monitoring [48–50]. Challenges include lower spatial resolution relative to other tomography methods and sensitivity to electrode placement, boundary conditions, and noise [48], which can be mitigated by increasing electrode count, optimising distribution, varying current injection schemes, or improving reconstruction algorithms [32,51–55].

Here, we demonstrate for the first time that lattice geometry can be deliberately engineered to systematically enhance EIT sensitivity and spatial localisation, effectively transforming the architected structure into an active component of the imaging system. This approach enables reliable detection of distant damage, sequential fracture mapping at distinct strain levels, and sensitive pre-failure detection. Building on this, we apply the method to 2D auxetic lattices fabricated from CNT-infused photocurable resins, integrating EIT with conventional piezoresistive characterisation to assess performance.

We focus on 2D auxetic lattices designed via Voronoi partitioning with branch–trunk–branch motifs inspired by 2D wallpaper symmetries. Mechanical and piezoresistive responses are first evaluated using end-to-end resistance measurements, followed by online EIT measurements during uniaxial tensile loading with adjacent and across current injection schemes. Reconstructed conductivity maps reveal localised ligament fractures and progressive failure sequences with high temporal resolution, while complementary global resistance measurements track macroscopic stress–strain evolution. Tunable aspects of lattice design further enable sensitive, spatially resolved detection of early-stage damage, emphasising EIT's advantage over conventional pointwise piezoresistive sensing. Collectively, these results establish *in situ* EIT as a robust, full-field modality for autonomous monitoring of multifunctional architected lattice materials, with potential applications in biomedical and structural health monitoring, as well as multifunctional energy-relevant systems. By integrating lattice design with advanced sensing, this work provides a new paradigm for simultaneous structural and functional characterisation in 3D-printed architected materials.

## 2. Electrical Impedance Tomography

Electrical impedance tomography (EIT) is a spatial imaging technique that reconstructs the distribution of electrical properties, such as conductivity and permittivity, within a material or body from boundary electrical measurements [29]. In self-sensing composites, these boundary measurements reflect changes in electrical impedance arising from strain or damage due to the inherent piezoresistive properties of the material, enabling full-field monitoring beyond discrete sensors. EIT involves solving both forward and inverse problems. The forward problem simulates boundary voltages for a known geometry (here, the lattice) given prescribed current injection, boundary conditions, and an assumed conductivity distribution [48]. This requires numerically solving the Laplace equation, which relates the conductivity distribution, $\sigma$, within a lattice domain, $\Omega$, with boundary $\partial\Omega$, to the potential field, $\phi$:

$$\nabla \cdot (\sigma \nabla \phi) = 0 \tag{1}$$

This equation is solved under complete electrode boundary conditions using the finite element method (FEM) [47]. Given the conductivity distribution, current injection scheme, measurement patterns, and domain geometry, FEM calculates the boundary voltages. A finite element solution for the 2D lattice domain is shown in **Figure S1** (Supplementary Information).

The inverse problem seeks to reconstruct the unknown conductivity distribution of the lattice from measured boundary voltages. It is inherently ill-posed and nonlinear [56] with additional challenges arising from electrode placement errors, system noise, and modelling uncertainties. To address these, we employ a dynamic imaging approach, estimating conductivity changes, $\Delta\sigma$, between sequential loading events using a Maximum a Posteriori (MAP) reconstruction framework [57]:

$$\Delta\sigma = (H^T W H + \mu Q)^{-1} H^T W \Delta V \tag{2}$$

Here, $H$ is the sensitivity (Jacobian) matrix encoding electrode positions, domain boundaries, and current injection/measurement patterns; $W$ is the noise covariance matrix, with diagonal entries representing the inverse variance of measurement noise; $Q$ is a Laplace prior (second-order high-pass filter) used for regularisation; and $\mu$ is the regularisation hyperparameter determined based on noise levels. This framework allows systematic, spatially resolved tracking of conductivity changes in the lattice during quasi-static tensile loading, providing a pathway to detect early-stage damage and map sequential fracture events. Further details regarding forward and inverse modelling are provided in Section S1 (Supplementary Information).

All forward FEM simulations, MAP implementation, and image reconstructions were performed using EIDORS, an open-source MATLAB (R2024a, MathWorks, USA)-based suite for electrical impedance and diffuse optical reconstruction [56]. To generate the finite element mesh, the 2D lattice geometry was first imported and meshed using MATLAB's PDE solver. The resulting mesh was then passed to EIDORS to define the FEM lattice domain. For comparison, a bulk rectangular domain with uniform conductivity was also considered to evaluate the influence of lattice architecture on image reconstruction. The MAP regularisation hyperparameter was determined based on the domain geometry for each case.

## 3. Materials and Methods

### 3.1. Geometric modelling of architected lattices

To systematically explore the influence of lattice geometry on multifunctional and self-sensing performance, we developed a computational geometric modelling framework for 2D architected lattices based on a Voronoi-partitioned branch–trunk–branch motif. This framework allows controlled tuning of lattice architecture, including branch configuration, unit cell density, and tiling, enabling a broad exploration of the lattice design space relevant for EIT-based full-field sensing.

The geometric modelling begins with a stem, comprising a trunk and six branches: two at each end and two at the midpoint of the trunk. The complete set of points in the stem is defined as:

$$S = T \cup \bigcup_{k=1}^{6} B_k \qquad (3)$$

where $T$ is the trunk, consisting of $n$ points, and $B_k = \{\boldsymbol{B_1}, \boldsymbol{B_2}, \boldsymbol{B_3}, \ldots \ldots \boldsymbol{B_m}\}$ is the set of $m$ points in the $k$-th branch. $S$ represents the complete stem, including the trunk and branches as shown in **Figure 1(a)**. The stem is arranged in the $x - y$ plane according to the wallpaper group $p4$, which applies rotations and translations to create a 4×4 tiling (**Figure 1(b)**). Each translated and rotated stem is given by:

$$S_{i,j} = \boldsymbol{R}_{\theta_{i,j}} S + i\boldsymbol{T_x} + j\boldsymbol{T_y} \qquad (4)$$

where $\boldsymbol{T_x} = (L, 0)$ and $\boldsymbol{T_y} = (L, 0)$ are the translation vectors,

$$\mathbf{R}_{\theta_{i,j}} = \begin{bmatrix} \cos\theta_{i,j} & -\sin\theta_{i,j} \\ \sin\theta_{i,j} & \cos\theta_{i,j} \end{bmatrix} \quad (5)$$

is the rotation matrix, and $i, j \in \{0,1,2,3\}$ index the tiling of 16 stems. All stems are combined into a single domain:

$$D = \bigcup_{i,j} S_{i,j} \quad (6)$$

The Voronoi diagram is then computed for all points $\mathbf{P} \in D$. The Voronoi cell, C for a seed point $\mathbf{P}$ is defined as:

$$C(\mathbf{P}) = \{\mathbf{Q} \in R^2 \mid d(\mathbf{Q}, \mathbf{P}) \leq d(\mathbf{Q}, \mathbf{P}'), \forall \mathbf{P}' \neq \mathbf{P}\} \quad (7)$$

where $d(\mathbf{Q}, \mathbf{P})$ is the Euclidean distance between points $\mathbf{Q}$ and $\mathbf{P}$. These Voronoi cells are used to construct unit cells, which are then offset, combined via Boolean operations, and extruded to produce the final 3D lattice geometry.

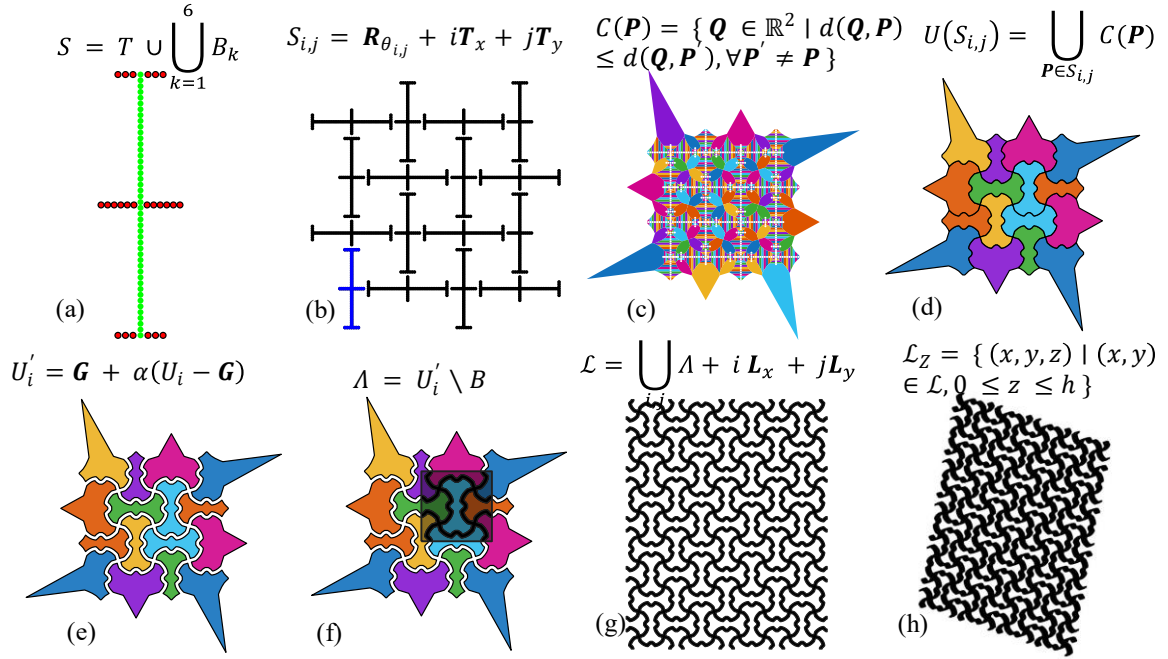

**Figure 1**: Mathematical framework for the lattice design. (a) Defining the stem; (b) arranging stems in a $p4$ symmetry pattern; (c) Voronoi decomposition of all points; (d) union of corresponding Voronoi cells; (e) contraction (offset, $\alpha$) of each polygon; (f) subtraction of the unit square $Q$ from polygons; (g) tiling the unit cell; (h) extrusion to the desired thickness.

For each stem $S_{i,j}$, the union of the Voronoi cells of all its points is computed:

$$U(S_{i,j}) = \bigcup_{\mathbf{P} \in S_{i,j}} C(\mathbf{P}) \quad (8)$$

resulting in 16 polygons (one per stem) as shown in **Figure 1(d)**. Each polygon $U_i$ is contracted inward by a scalar α, determined by the target relative density of the lattice:

$$U_i' = G + \alpha(U_i - G) \tag{9}$$

where $G$ is the centroid of the polygon $U_i$ and is defined as

$$G = \frac{1}{|U_i|} \sum_{P \in U_i} P \tag{10}$$

The unit cell $\Lambda$ is obtained by subtracting a square $B$ (of unit cell dimensions) from the union of the 16 contracted polygons (**Figure 1(f)**):

$$\Lambda = U_i' \setminus B \tag{11}$$

The final lattice geometry $\mathcal{L}$ is constructed by tiling the unit cell in $x$ and $y$ directions:

$$\mathcal{L} = \bigcup_{i,j} (\Lambda + i\, L_x + j\, L_y) \tag{12}$$

where $L_x$ and $L_y$ are the unit cell translation vectors along $x$ and $y$. The 2D lattice is then extruded in the $z$-direction to obtain the 3D metamaterial $\mathcal{L}_z$:

$$\mathcal{L}_Z = \{(x, y, z) \mid (x, y) \in \mathcal{L}, 0 \leq z \leq h\} \tag{13}$$

with extrusion height $h = 4$mm as shown in **Figure 1(h)**. Each unit cell measures 12 mm ×12 mm, arranged in a 4×5 grid along $x$ and $y$. The complete modelling framework, from generation of a single stem to the formation of the 2D lattice is demonstrated in **Supplementary Video SV1**. The lattices were modelled in MATLAB, and four distinct lattice geometries were explored in this study (**Table S1**, Supplementary Information). By systematically varying these geometric parameters, the framework allows controlled exploration of lattice design for both mechanical performance and EIT-based sensing sensitivity.

### 3.2. Fabrication of metamaterials

Architected lattice composites were fabricated using a DLP 3D printer (Asiga Pro 4K 65UV, 385 nm), enabling high-resolution control over geometry and material distribution. A photosensitive resin (PlasGray) was rendered electrically conductive by incorporating multiwalled carbon nanotubes (MWCNTs), forming a printable nanocomposite suitable for *in situ* EIT. To prepare the nanocomposite resin, 2.43 wt.% triphenylphosphine oxide (TPO) was first dissolved in 48.75 wt.% tripropylene glycol diacrylate (TPGDA) and magnetically stirred for 20 min to ensure complete mixing. Subsequently, 0.049 wt.% MWCNTs were added and dispersed via probe sonication for 8 min at low amplitude, yielding a homogeneous suspension without visible agglomerates. Finally, 48.75 wt.% PlasGray was added, followed by 1 h of magnetic stirring to achieve a stable, uniform nanocomposite resin. This formulation follows the approach reported by Johannes et al. [58].

Printing parameters were systematically optimised to achieve high mechanical fidelity and interlayer uniformity. Layer thickness, light intensity, and exposure time were set to 50 μm, 7 mW·cm$^{-2}$, and 5.46 s, respectively. Three types of specimens were fabricated: (i) dogbone samples (ASTM D638-14) for tensile characterisation, (ii) pristine architected lattices, and (iii) pre-cracked lattices containing two intentionally introduced notches on ligaments adjacent to the central unit cell. The pre-cracked lattices enabled assessment of EIT's ability to detect fractures remote from electrode locations (Supplementary Information, **Figure S2**). Following printing, all specimens were washed in isopropyl alcohol to remove uncured resin and post-cured under UV light for 6 min. The resulting nanocomposite lattices exhibited excellent dimensional accuracy, structural integrity, and surface finish, suitable for reliable electrical and mechanical testing.

For EIT electrode integration, sixteen equally spaced electrodes were positioned along the lattice periphery. Silver conductive paint (RS Pro) was applied at each contact region, and multistrand copper wires were attached using a flexible silver epoxy adhesive (Flexible-Silver 16, Epoxy International). A detailed electrode attachment procedure is provided in Section S2 and **Figure S3** of the Supplementary Information. The compliant nature of the silver epoxy accommodated local deformations during loading while maintaining stable, low-resistance electrical contact (Supplementary Information, **Figure S4**). This robust and reproducible electrode–lattice interface ensured low and temporally stable contact impedance, enabling high-fidelity signal acquisition under *in situ* mechanical loading. The same electrode preparation protocol was applied to both bulk and lattice specimens employed for direct resistance measurements.

Overall, this fabrication route establishes a scalable and reproducible strategy for producing conductive architected lattices with precisely defined microarchitecture and robust electromechanical coupling. The method provides a solid experimental foundation for real-time, spatially resolved EIT-based damage sensing, enabling high-quality reconstruction of evolving conductivity fields in multifunctional metamaterials.

**3.3 Sensitivity quantification for EIT-based damage detection**

The sensitivity of EIT measurements to local conductivity perturbations depends on the spatial distribution of current density within the lattice, which is inherently non-uniform [59].

Conductivity changes occurring in regions of low current density may induce voltage variations below the noise floor, motivating the need for architectures exhibiting high minimum sensitivity throughout the domain.

To quantify the spatial sensitivity of the EIT configuration, a Jacobian-based analysis was performed. Specifically, the $L^2$-norm of each column of the Jacobian matrix ($H$), representing the sensitivity of each voltage measurement to local conductivity changes, was computed and normalised by the total number of voltage measurements [60]:

$$\text{Sensitivity} = \frac{1}{N}[\|h_1\|_2, \|h_2\|_2 \ldots] \tag{16}$$

where $h_i$ denotes the $i$-th column of the Jacobian matrix, $H$ and $N$ is the number of measurements for a given current injection–measurement protocol (208 for the adjacent injection scheme). This metric provides a consistent, geometry-independent basis for comparing EIT sensitivity across different lattice architectures and loading configurations.

### 3.4. Experimental testing

### 3.4.1. Mechanical and piezoresistive response

Uniaxial tensile tests were performed on both bulk and lattice composite specimens using a universal testing machine (Zwick Roell Z250, 250 kN load cell). All tests were conducted under quasi-static conditions at a constant crosshead displacement rate of 1 mm min$^{-1}$. Full-field strain measurements for the bulk composites were obtained using three-dimensional digital image correlation (3D-DIC, VIC-3D, Correlated Solutions, Inc., USA) with a subset size of 29 and a step size of 9. The imaging system employed dual CSI-acA2440-75 µm cameras equipped with Xenoplan 2.0/28-0901 lenses. Each experiment was repeated three times to ensure reproducibility and statistical validity.

The piezoresistive response of both bulk and lattice architectures was characterised concurrently with mechanical testing by monitoring the electrical resistance using a high-precision multimeter (Fluke 8846A, measuring range: $10\,\Omega - 1\,\text{G}\Omega$, resolution: $10\,\mu\Omega$). The global strain sensitivity was quantified in terms of the gauge factor ($GF$), defined as

$$GF = \frac{((R - R_0)/R_0)}{\varepsilon} \tag{14}$$

where $R_0$ and $R$ are the electrical resistances of the specimen in the unstrained and strained states, respectively, and $\varepsilon$ is the applied strain. The gauge factor provides a direct measure of piezoresistive sensitivity, quantitatively capturing the electromechanical coupling in the AM-fabricated architected composites.

### 3.4.2. *In situ* electrical impedance tomography

*In situ* EIT measurements were performed concurrently with quasi-static tensile testing to monitor the spatial evolution of conductivity within the architected lattice. Two current injection configurations—adjacent and across—were implemented, while voltage measurements followed an adjacent protocol in both cases.

In the adjacent current injection scheme, current was sequentially injected between neighbouring electrode pairs (1–2, 2–3, 3–4, …), while voltages were recorded across all remaining adjacent electrode pairs. This procedure was repeated until each of the sixteen electrodes had served as a current source, resulting in 208 voltage measurements per acquisition cycle ( 16 injections × 13 measurements). **Figure 2a** illustrates the spatial distribution of current flow, where arrow orientation indicates current direction and colour represents local voltage magnitude.

In the across current injection scheme, current flowed between geometrically opposite electrode pairs (1–12, 2–11, 3–10, …) while retaining the adjacent voltage measurement protocol (**Figure 2b**). This configuration generated 192 measurements per cycle ( 16 injections × 12 measurements). In both schemes, voltage measurements were omitted between electrodes actively engaged in current injection to avoid the influence of contact impedance [42].

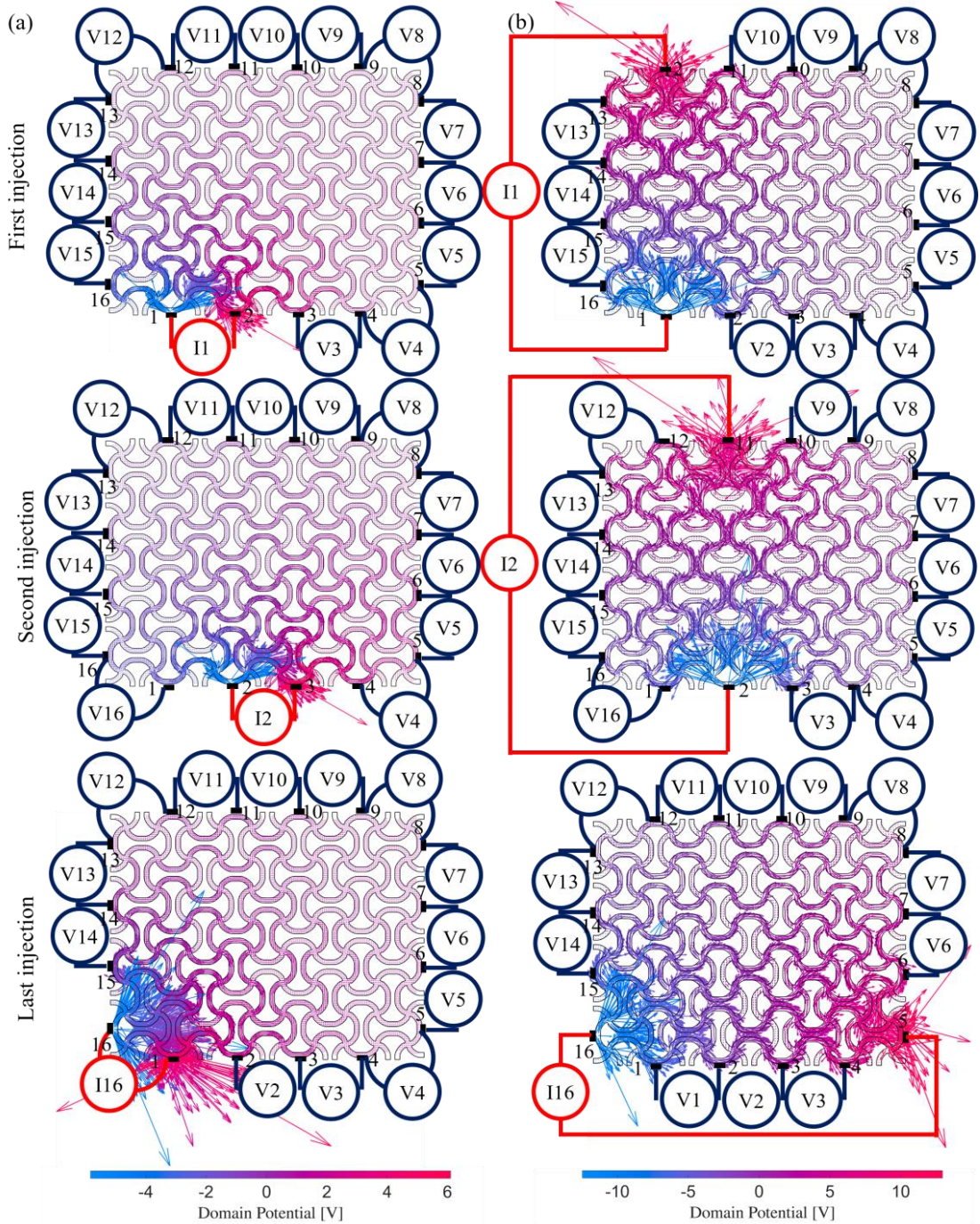

**Figure 2:** Schematic showing (a) adjacent and (b) across current injection with adjacent voltage measurements.

A machine displacement rate of 0.1 mm min⁻¹ was used for all EIT-integrated tests to ensure reliable data acquisition and allow sufficient time to complete each high-density measurement cycle. Data were acquired using the Spectro-EIT system[61], developed in collaboration with Eliko (Estonia), Tallinn University of Technology (Estonia), and Tampere University (Finland). This system provides multi-frequency excitation (1 − 349 kHz across 15 frequencies) with an embedded multiplexer, enabling rapid, high-fidelity impedance mapping. Both resistance and voltage were measured. For the present study, measurements were conducted at 1 kHz, corresponding to cycle times of approximately 3.5 s and 3.3 s for the adjacent and across injection schemes, respectively. Voltage data were streamed to a MATLAB-based EIDORS reconstruction pipeline for real-time conductivity mapping (**Figure 3a**).

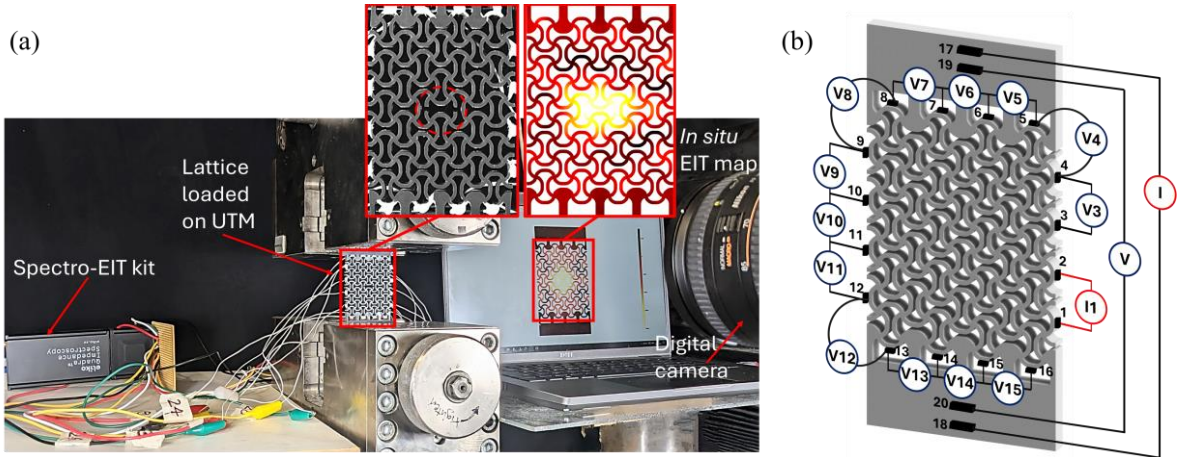

**Figure 3:** (a) Experimental setup for *in situ* EIT showing real-time conductivity reconstruction. (b) Schematic illustrating simultaneous end-to-end resistance measurement alongside EIT.

To complement EIT-based spatial mapping, Four-Point Probe (4PP) measurements were performed simultaneously along the longitudinal loading direction to monitor global resistance evolution (**Figure 3b**). The 4PP configuration effectively eliminated errors from fluctuating contact resistance, ensuring accurate global resistance quantification [62]. Together, these methods provided a dual-scale sensing framework: EIT captured localised conductivity variations, while 4PP quantified the overall electromechanical response. In total, twenty electrodes were employed: sixteen for EIT and four for 4PP. The Spectro-EIT device accommodates up to thirty-two electrodes, enabling synchronised data acquisition. Each complete EIT cycle (208 measurements) was immediately followed by a global resistance measurement. All experiments were recorded using a high-resolution digital camera for temporal correlation between mechanical deformation and conductivity evolution.

For quantitative analysis of EIT tomograms, a matrix-based approach was used to identify the element exhibiting the maximum negative conductivity change $\Delta\sigma_{min}$ at each deformation state. The magnitude of this change, $|\Delta\sigma_{min}|$, served as a scalar indicator representing the most responsive zone within the lattice. This metric provides a consistent and physically meaningful measure of local conductivity evolution, enabling direct comparison across deformation stages and specimens.

To evaluate the aggregate piezoresistive response, an effective gauge factor ($GF_e$) was computed. All 208 directly measured resistances from the EIT system were summed at each strain state. Similar to conventional gauge factor calculations, $GF_e$ was obtained from the slope of cumulative relative resistance change versus global strain ($\varepsilon$):

$$GF_e = \frac{\sum (R-R_0)/R_0}{\varepsilon} \tag{15}$$

## 4. Results and Discussion

### 4.1 Sensitivity analysis

**Figure 4** presents the spatial sensitivity heat maps and corresponding violin plots for four lattice designs randomly selected from the design space. Lattice **A** exhibits the highest mean sensitivity ($2.37 \times 10^{-6}$ V·m·S$^{-1}$), followed by lattice **C** ($2.23 \times 10^{-6}$ V·m·S$^{-1}$), lattice **D** ($2.03 \times 10^{-6}$ V·m·S$^{-1}$), and lattice **B** ($1.58 \times 10^{-6}$ V·m·S$^{-1}$). Notably, the 25th-percentile sensitivity of lattice **A** exceeds that of all other designs, indicating higher minimum sensitivity across the domain. Based on this analysis, all subsequent experiments were performed on lattice **A**.

The geometric design of the lattices directly governs EIT performance. By tuning branch lengths, junction angles, and the symmetry of the branch–stem–branch motif, lattice topology amplifies spatial variations in conductivity under mechanical loading. Voronoi-based unit cells with controlled polygonal contraction (α) enable precise modulation of relative density, thereby tailoring local current pathways and electrode voltage responses. The imposed p4 symmetry introduces periodicity that ensures a predictable correspondence between electrode locations and lattice features, which is essential for high-fidelity reconstruction. Consequently, the lattice functions not only as a load-bearing structure but also as an active contributor to EIT imaging, enabling robust and spatially resolved sensing.

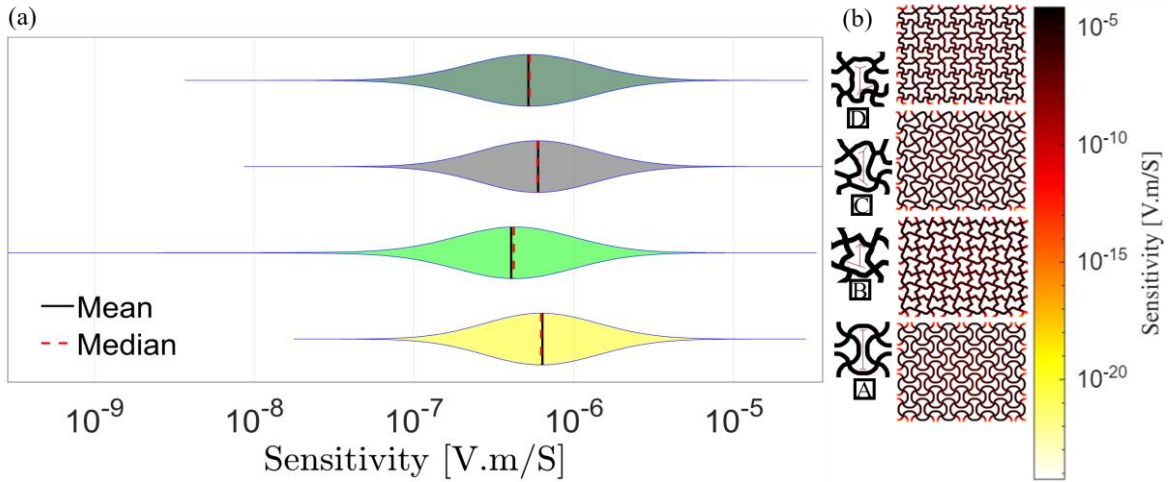

**Figure 4:** (a) Violin plots and (b) spatial maps of element-wise sensitivity (log scale) from the forward model for four lattice architectures, highlighting the influence of geometry on EIT detectability and uniformity of sensitivity.

### 4.2. Scanning electron microscopy

To evaluate the dispersion of multi-walled carbon nanotubes (MWCNTs) within the PlasGray matrix, cryogenically fractured samples were prepared by immersing them in liquid nitrogen immediately prior to fracture. The fractured surfaces were sputter-coated with a thin gold layer to enhance conductivity and mitigate charging during imaging. SEM analysis (TESCAN CLARA) revealed that the MWCNTs are uniformly distributed throughout the polymer matrix (**Figure 5**), ensuring effective electrical percolation and consistent conductive pathways. This uniform dispersion is critical for achieving reliable piezoresistive behavior and high-fidelity EIT performance in the architected lattices.

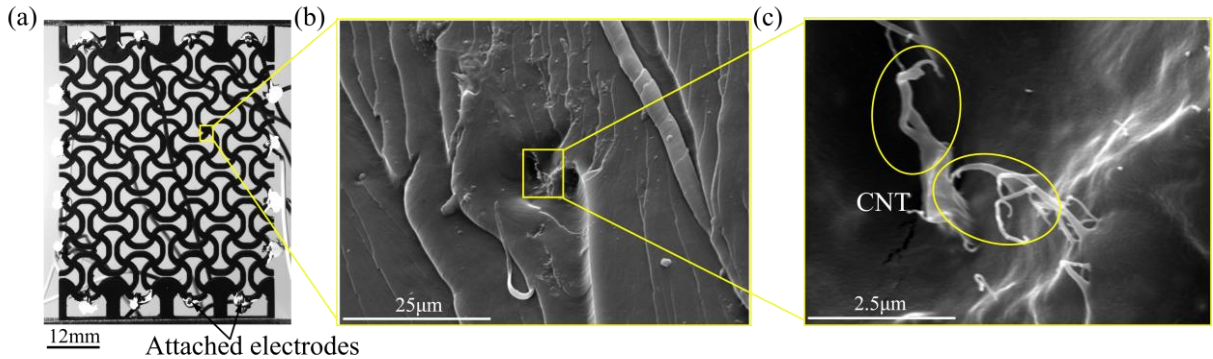

**Figure 5:** (a) Intact 2D lattice with electrodes positioned along the periphery, (b, c) SEM images of the cryogenically fractured PG/CNT composite showing uniform MWCNTs dispersion.

### 4.3. Mechanical and piezoresistive performance of parent and lattice composites

The mechanical response of the parent PG/CNT composites under tensile loading is provided in the Supplementary Information (**Figure S5**). The PG/CNT composite exhibits a brittle response, with tensile yield and ultimate strengths of $19.88 \pm 1.07$ MPa and $31.48 \pm 0.37$ MPa, respectively, and a Young's modulus of $1392.6 \pm 13.6$ MPa. DIC analysis yielded a Poisson's ratio of 0.403 within the elastic regime, a yield strain of $0.0163 \pm 0.0008$, and a failure strain of $0.0716 \pm 0.0079$. The electrical resistance increased monotonically with applied strain, as tensile deformation induces separation of the CNT conductive network, increasing interparticle tunnelling distances and thus electrical resistance [26]. The corresponding gauge factor, quantifying the electromechanical coupling in the parent composite, was measured to be approximately 4.2.

The in-plane uniaxial response of the autonomous sensing PG/CNT 2D lattice was evaluated under quasi-static tensile loading. **Figure 6** presents the macroscopic stress–strain and piezoresistive responses, together with deformation and failure maps at selected strain levels. **Supplementary Video SV2** shows synchronized stress-strain and piezoresistive responses, along with corresponding deformation maps, obtained from experiments. The lattice exhibits an initial linear-elastic regime followed by progressive work hardening up to ligament fracture, which triggers rapid, catastrophic failure of the lattice. This failure mode is governed by the intrinsic brittleness of the parent composite; however, the architected lattice topology effectively redistributes deformation, resulting in enhanced macroscopic strain tolerance relative to the bulk material. The Young's modulus and tensile strength of the lattice were measured as 33.44 ± 0.3689 MPa and 3.63 ± 0.26 MPa, respectively, with a failure strain of 0.1069 ± 0.0124, highlighting the ability of the lattice architecture to accommodate larger strains despite the brittle nature of the constituent material.

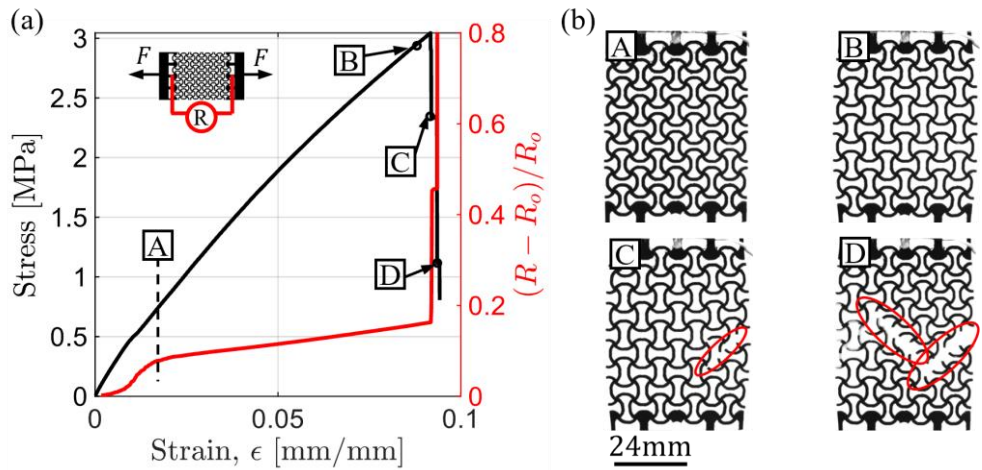

**Figure 6:** (a) Stress–strain and piezoresistive responses of the PG/CNT lattice under in-plane quasi-static tensile loading. (b) Deformation and failure maps at different strain levels. The correlated evolution of mechanical response and resistance change reveals the progressive ligament fracture and associated disruption of the conductive network within the architected lattice.

The piezoresistive response was quantified through the fractional change in electrical resistance as a function of applied macroscopic strain. As shown in **Figure 6a**, the electrical resistance increases monotonically with tensile strain, primarily due to microstructural stretching of the CNT-based conductive networks within the ligaments. Tensile deformation increases interparticle tunnelling distances and progressively disrupts conductive pathways, leading to a rise in global resistance. A pronounced increase in resistance coincides with the onset of ligament fracture, followed by a steep rise associated with rapid fracture propagation and conductive network breakdown. The gauge factor ($GF$) of the lattice was determined to be 1.02, corresponding to 24.29% of the bulk composite's GF, reflecting the trade-off between sensitivity and enhanced strain accommodation introduced by the lattice architecture while maintaining a stable and damage-correlated sensing response.

### 4.3. *In situ* EIT-enabled damage sensing in architected lattice composites

#### 4.3.1. Pristine architected lattice: *in situ* EIT mapping of deformation and fracture

The macroscopic stress–strain response, cumulative relative resistance changes as a function of strain, and the corresponding *in situ* EIT reconstructions for the PG/CNT lattice composite are shown in **Figure 7**. Five representative deformation states (**A**–**E**) were selected to capture the progressive evolution of deformation and damage during tensile loading. Image reconstructions were performed using the adjacent current injection and voltage measurement scheme. For consistency across deformation states, the conductivity amplitude in each EIT map was normalised by the maximum amplitude recorded at the near-fracture state, yielding normalised conductivity values within the range [0,1] and presented on a logarithmic scale (**Figure S7a**). In addition, conductivity changes in each reconstructed image were normalised within their respective frames to highlight localised damage evolution (**Figure 7c**, middle row). **Supplementary Video SV3** shows the synchronised stress–strain response, cumulative relative resistance changes with respect to macroscopic strain, logarithmic image amplitude alongside EIT-reconstructed video, and uniaxial mechanical testing footage.

At the initial deformation stage (state **A**), a localised reduction in conductivity—corresponding to an increase in electrical resistance—was detected near the region that subsequently developed into the primary fracture site. Upon ligament rupture (state **B**), a sharp and spatially confined increase in local resistance was observed, precisely coinciding with the fracture location, demonstrating the high spatial sensitivity of the EIT framework. With continued loading (states **C**–**E**), successive ligament fractures were distinctly resolved, each associated with a pronounced drop in local conductivity. The magnitude of these conductivity changes increased systematically with applied strain, reflecting progressive damage accumulation and conductive network degradation within the lattice. Comparable trends were observed in images reconstructed over the bulk domain (**Figure 7c**, bottom row), confirming the robustness and geometric independence of the EIT-based damage detection, as the localised conductivity loss consistently overlapped with the physically observed fracture sites. Collectively, these results demonstrate the ability of *in situ* EIT to capture both the spatial and temporal evolution of damage in architected lattice composites in real time.

To quantify the global electromechanical response, all 208 directly measured resistances from the EIT system were summed at each deformation state, and the relative change in the summed resistance was plotted as a function of applied strain (**Figure 7b**, bottom left panel). The resulting response exhibits a monotonic and approximately linear increase with strain, corresponding to an effective gauge factor of 0.64, which is 62.74% of the gauge factor obtained from conventional two-probe resistance measurements. Notably, the fracture of individual ligaments or struts produces abrupt increases in the summed resistance due to the loss of dominant conductive pathways, providing a clear electrical signature of discrete damage events.

In the conventional resistance measurement approach, the electrical response is primarily measured along the loading direction, thereby maximising sensitivity to uniaxial deformation. In contrast, the summed EIT resistance represents an effective global response derived from multiple current paths spanning different orientations within the lattice. Consequently, the EIT-based gauge factor is reduced but reflects a spatially averaged, more robust electromechanical response, inherently less sensitive to localised measurement artefacts. This trade-off between sensitivity and robustness underscores the suitability of EIT for reliable, distributed damage sensing in architected lattice composites.

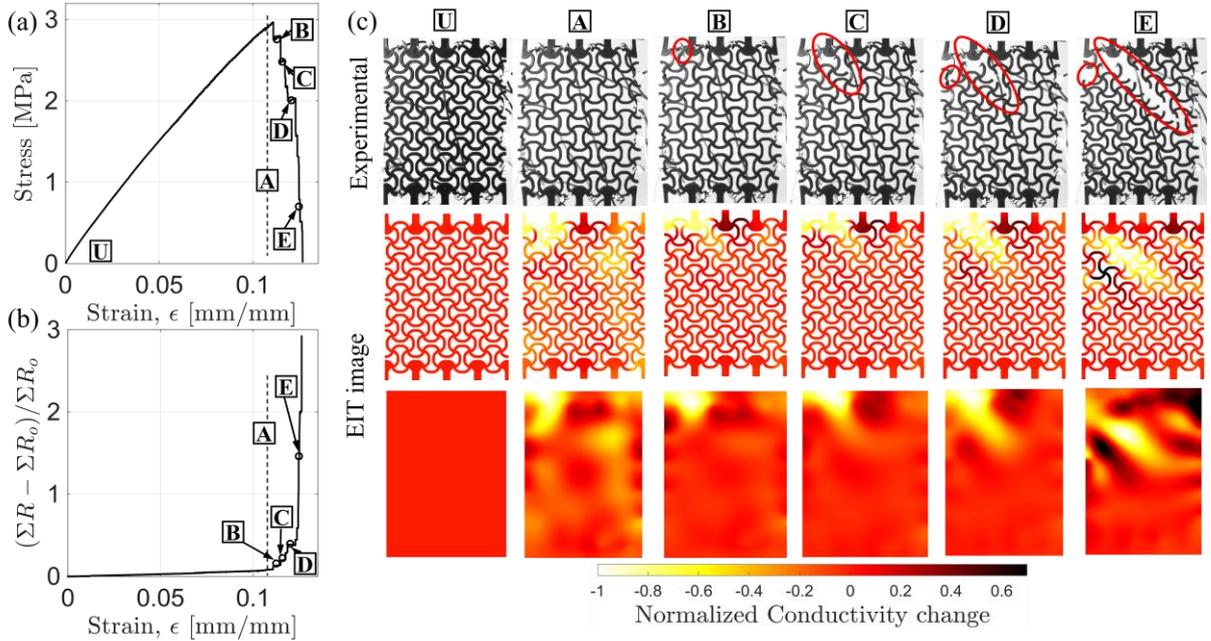

**Figure 7:** (a) Macroscopic stress–strain response, (b) cumulative relative resistance change as a function of strain, and (c) deformation, failure maps, and *in situ* reconstructed conductivity images of the PG/CNT lattice under tensile loading.

The coupling between mechanical deformation and electrical response is further elucidated in **Figure 8**, which shows the distribution of all 208 measured resistance changes relative to the current injection electrodes for deformation stages **A**-**E**. The resistance changes were calculated with respect to the undeformed state **U**. A pronounced increase in measured resistances was observed in the regions adjacent to the fractured ligaments, whereas regions remote from the damage exhibit comparatively minor changes. With progressive ligament fractures (stages **D** and **E**), the resistance increased near electrodes 1-4, consistent with the proximity of the advancing fracture propagation path to these electrodes. This spatial contrast directly reflects the localised disruption of conductive pathways and substantiates the high localisation fidelity and sensitivity of the EIT framework to microscale damage events within the architected lattice. **Figure S8** presents the distribution of all 208 measured resistances for three representative states: the unstrained configuration (violet), the post-fracture state following ligament failure between electrodes 7 and 8 (green), and the corresponding resistance change between these two states (red), along with the corresponding EIT image.

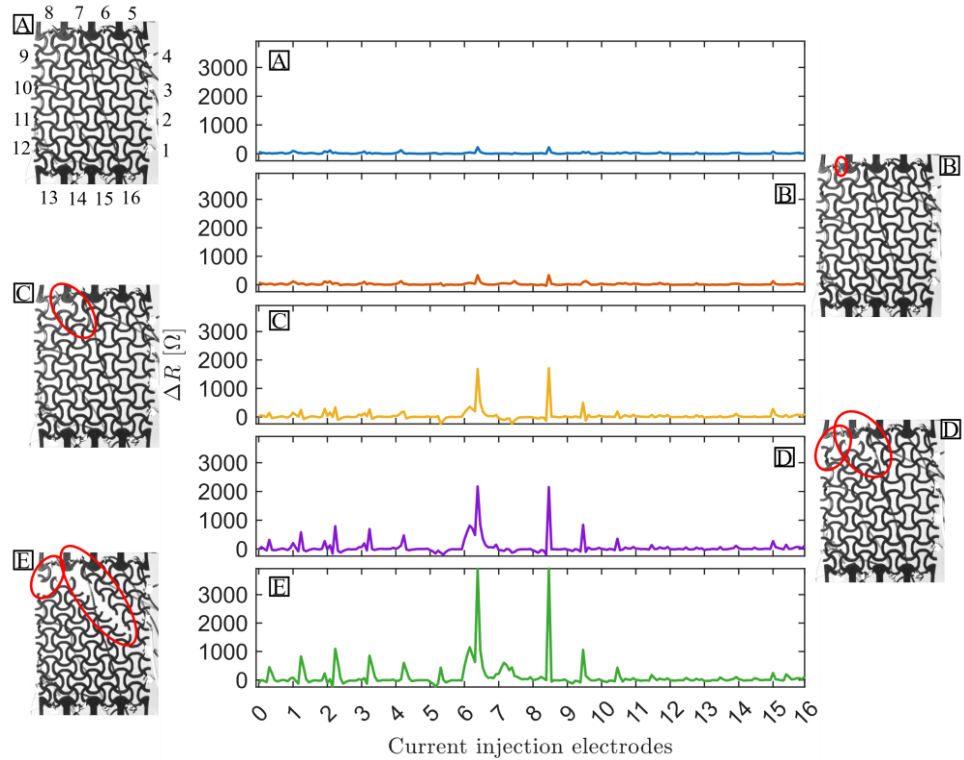

**Figure 8:** Distribution of the measured resistances for all 16 current injection configurations in the intact specimen at different strain states (**A-E**).

The evolution of the minimum absolute conductivity change, $|\Delta\sigma|_{min}$, as a function of applied strain is shown in **Figure S9a**. The magnitude of $|\Delta\sigma|_{min}$ increases monotonically with increasing strain and exhibits a distinct jump at the onset of fracture, consistent with the abrupt loss of conductive pathways during ligament failure. Similar trends were observed in the logarithmic scaled image amplitude of the lattice as shown in **Figure S8a**. This matrix-based quantification of conductivity evolution shows strong correspondence with both the macroscopic stress–strain response and the cumulative resistance trends, thereby validating the robustness, sensitivity, and interpretive accuracy of the *in situ* EIT approach for tracking damage initiation and progression in architected lattice composites.

**4.3.2. Pre-cracked architected lattice: EIT localisation of remote damage**

To evaluate the capability of EIT to localise damage occurring remotely from electrode positions, *in situ* EIT measurements were performed on a 2D lattice specimen containing a centrally located pre-crack. The macroscopic stress–strain response, cumulative relative resistance change as a function of strain, and the corresponding deformation and reconstructed conductivity maps are presented in **Figure 9**. Importantly, individual ligament fracture events— including those occurring within the lattice interior and away from the electrode periphery—were distinctly captured in the EIT reconstructions, demonstrating the ability of the technique to resolve spatially distributed damage in architected lattices.

Regions of pronounced conductivity loss in the reconstructed images coincided precisely with the experimentally observed fracture locations, and their spatial evolution closely tracked successive ligament failures emanating from the pre-crack. The cumulative relative resistance change increased steadily with applied strain and exhibited sharp, discrete spikes associated with abrupt ligament fracture events, highlighting strong electromechanical coupling and the high temporal sensitivity of the EIT framework. Consistent trends were also observed in the evolution of the minimum absolute conductivity change, $|\Delta\sigma|_{min}$, with applied strain (**Figure S7b**), further corroborating the robustness of the EIT-based damage quantification. The effective gauge factor for the pre-cracked lattice was determined to be 0.602.

In addition, the mechanical and simple electrical resistance change response was evaluated during an in-plane uniaxial tensile test of the pre-cracked 2D lattice. The correspond mechanical and piezoresistive response as a function of strain along with deformation maps in case of pre-crack lattice is shown in **Figure S9**. The synchronised experimental footage is provided in **Supplementary Video SV4**.

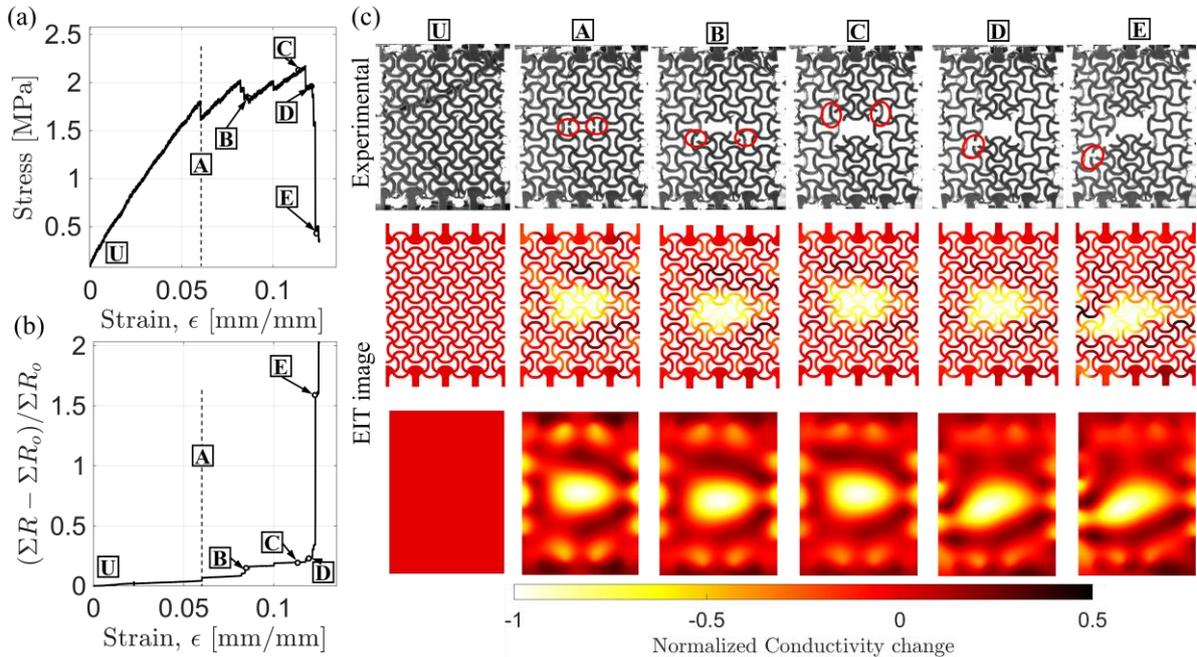

**Figure 9:** Pre-cracked lattice composite under tensile loading: (a) macroscopic stress–strain response, (b) cumulative relative resistance change as a function of strain, and (c) deformation maps and corresponding *in situ* EIT reconstructions highlighting spatially resolved fracture initiation and progression.

To further assess the robustness and geometric generality of the EIT approach, *in situ* measurements were additionally performed on pre-cracked specimens using an across-current injection and adjacent-voltage measurement scheme. This configuration promotes a more homogeneous sensitivity matrix and enhances the detection of conductivity changes arising from regions distant from the electrode boundaries [63]. The corresponding macroscopic stress–strain response, deformation maps, and conductivity evolution are provided in the Supporting Information (**Figure S11**), confirming the improved spatial uniformity and reliability of remote fracture detection achieved using the across-injection scheme.

### 4.3.3 Benchmarking *in situ* EIT against end-to-end four-point probe measurements

To directly benchmark *in situ* EIT against conventional electrical resistance measurements, simultaneous end-to-end four-point probe (4PP) resistance measurements and EIT imaging were performed on the 3D-printed PG/CNT lattice composite. The evolution of global resistance, cumulative relative resistance change, and end-to-end relative resistance change as a function of applied strain, together with the corresponding deformation maps and reconstructed conductivity fields, is presented in **Figure 10**.

The 4PP resistance increased monotonically with applied strain, capturing the bulk electromechanical coupling of the lattice and yielding a global gauge factor. In parallel, the EIT reconstructions revealed spatially localised conductivity losses associated with individual ligament fracture events. These localised conductivity reductions coincided precisely with fracture initiation and propagation sites observed experimentally, demonstrating the ability of EIT to image distributed damage evolution in real time. In contrast, both the cumulative relative resistance change and the end-to-end 4PP resistance represent spatially averaged electrical responses of the lattice and provide no information on damage localisation or strain heterogeneity, as shown in **Figure 10b**. For the specimen investigated, an effective gauge factor of 0.7181 was obtained.

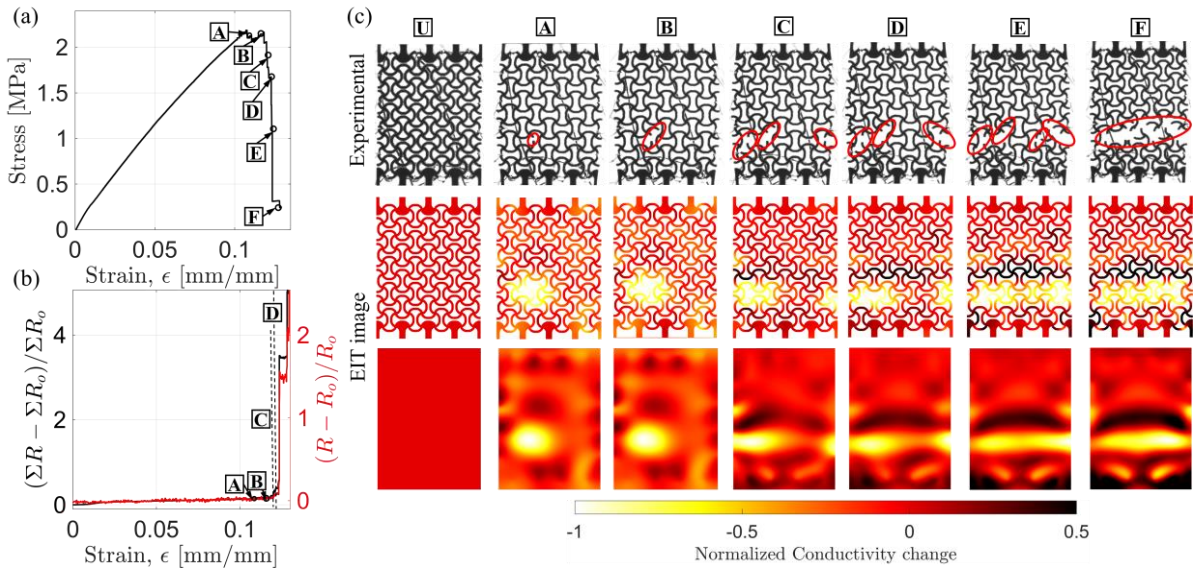

**Figure 10:** Simultaneous end-to-end electrical and *in situ* EIT measurements under tensile loading: (a) macroscopic stress–strain response, (b) cumulative relative resistance change and global resistance evolution, and (c) deformation/failure maps and reconstructed conductivity fields highlighting spatially localised damage.

It is noteworthy that in our recent studies [26,64], a multiscale piezoresistive finite element framework was developed to numerically reconstruct strain-induced conductivity fields in additively manufactured lattice architectures. While this modelling approach successfully captured the spatial variation of piezoresistive response, it required extensive constitutive calibration, significant computational effort, and ultimately relied on experimental validation. Moreover, conventional piezoresistive measurement techniques are inherently limited to discrete measurement points or bulk-averaged responses and cannot directly provide spatially resolved resistance distributions. In contrast, EIT provides a measurement-driven, non-destructive experimental framework for reconstructing spatially resolved conductivity fields without the need for prior calibration or complex inverse electromechanical coupling models. As such, EIT effectively bridges the gap between global resistance monitoring and numerically reconstructed piezoresistive fields, enabling direct experimental visualisation of strain localisation and damage evolution within architected lattices.

This comparative assessment highlights the complementary strengths of the two approaches. Piezoresistive finite element modelling offers predictive capability and mechanistic insight into the multiscale coupling between deformation and electrical response, whereas EIT delivers real-time, experimentally derived spatial imaging of conductivity evolution, making it particularly suited for autonomous sensing and model validation. Together, these techniques establish a powerful hybrid framework for the design, validation, and deployment of multifunctional self-sensing lattice materials—combining EIT-based global tomography with simulation-informed local quantitative analysis. Future advances in EIT reconstruction, including improved regularisation strategies, optimised electrode configuration and adaptive meshing techniques [31], are expected to further enhance spatial resolution and integration with predictive modelling frameworks.

## 5. Conclusion

This study demonstrates, for the first time, the *in situ* application of electrical impedance tomography (EIT) for full-field, real-time monitoring of damage evolution in architected lattice composites. By integrating EIT within CNT-infused 3D-printed lattices designed using tuneable Voronoi-based branch–trunk–branch motifs inspired by 2D wallpaper symmetries, we systematically explored the interplay among lattice architecture, electromechanical response, and sensing performance. The reconstructed conductivity maps provide high-fidelity visualisation of sequential ligament fractures, capturing both localised damage near electrodes and remote events, including those originating from pre-existing defects, and enabling the detection of early-stage, pre-failure damage. Our results show that lattice geometry can be deliberately tuned to enhance EIT sensitivity, spatial uniformity, and resolution, effectively transforming the structure into an active component of the sensing system. Discrete damage events manifest as pronounced conductivity changes, while cumulative fracture progression is reflected in monotonic trends in global and local resistance. Comparisons with conventional piezoresistive and four-point probe measurements highlight the unique advantage of EIT in providing spatially continuous, non-invasive, and calibration-free imaging of internal damage. Beyond its experimental novelty, this framework introduces a new paradigm for multifunctional architected materials, enabling simultaneous structural and functional characterisation in a single measurement modality. By coupling lattice design, material composition, and electrical sensing, we provide a scalable platform for intelligent materials that can support digital twin frameworks, predictive maintenance strategies, and advanced structural health monitoring. The methodology is broadly applicable across multiple length scales, with immediate relevance to structural, biomedical, and energy systems where precise, real-time damage detection is critical.

Collectively, these findings establish EIT-enabled architected lattices as a transformative approach for autonomous, full-field sensing, offering both mechanistic insight and practical utility. By demonstrating that lattice geometry can be strategically engineered to maximise sensing fidelity and that electrical signals correlate robustly with progressive mechanical failure, this work lays the foundation for rational design of intelligent, multifunctional materials. Integration with advanced EIT reconstruction algorithms, optimised electrode configuration, adaptive meshing, and predictive multiscale modelling promises further improvements in spatial resolution, sensitivity, and real-time interpretability, paving the way toward next-generation self-monitoring materials that are both load-bearing and intrinsically intelligent.

**Acknowledgements**

The Authors would like to thank James Gallagher for his assistance with the SEM imaging. The authors would also like to acknowledge the financial support through the USyd-UofG Ignition Grants. S. K. gratefully acknowledges the support through the Vaibhav Fellowship awarded by the Indian National Academy of Engineering (INAE)/Department of Science & Technology, India [DST/IC/VAIBHAV/Award/2024/L-10].

Supporting Information

**Full-Field Damage Monitoring in Architected Lattices Using *In Situ* Electrical Impedance Tomography**

*Akash Deep, Andrea Samore, Alistair McEwan, Andrew McBride, and Shanmugam Kumar\**

A. Deep, S. Kumar

Sustainable Multifunctional Materials and Additive Manufacturing (SM$^2$AM) Laboratory,

James Watt School of Engineering, University of Glasgow, Glasgow G12 8QQ, UK

E-mail: Msv.Kumar@glasgow.ac.uk

A. Samore, A. McEwan

School of Biomedical Engineering, University of Sydney, Sydney, New South Wales,

Australia

A. McBride

Glasgow Computational Engineering Centre, James Watt School of Engineering, University

of Glasgow, Glasgow G12 8QQ, UK

**S1. Forward and inverse formulation of electrical impedance tomography (EIT)**

EIT aims to reconstruct the spatially distributed conductivity map of a domain, with electrodes at its periphery, given measured electrical signals. It requires solving an inverse problem that seeks to estimate the change in conductivity by minimising the differences between numerically computed and experimentally measured boundary voltages. Scalar quantities are denoted by italic lowercase symbols, vectors by bold italic lowercase symbols, matrices by bold italic uppercase symbols, and sets by calligraphic upright symbols. Modelling voltage distribution requires solving the Laplace equation, which governs the relationship between current and voltages within a domain Ω having a boundary, ∂Ω as shown in the equation below [1].

$$\nabla . \sigma \nabla \phi = 0 \qquad (1)$$

Here, σ is the spatially varying conductivity distribution and ϕ is the potential within the domain. This differential equation is subjected to the complete electrode model (CEM), which considers the potential drop across the electrodes due to contact impedance. [2]

$$\phi + z_l \sigma \nabla \phi . \boldsymbol{n} = V_l \qquad (2)$$

Here, $z_l$ is the contact impedance between the domain and the $l$th electrode, **n** is the outward pointing normal vector. The conservation of charge is also enforced using the equation below

$$\sum_{l=1}^{L} \int_{E_l} \sigma \nabla \phi \cdot \mathbf{n} dE_l = 0 \quad (3)$$

In this equation, $E_l$ represents the length of the $l$th electrode and $L$ is the total number of electrodes which, in our study, is sixteen. The Laplace equation, subjected to the above-stated boundary conditions, can be solved numerically using the finite element method. More information on the FE formulation is available in the literature. [1,2] Thus, for a given domain with known conductivity, injection and measurement patterns across electrodes with known electrical impedance and injected current, boundary voltages across the electrodes can be computed numerically.

In this study, the lattice geometry was meshed with 2D triangular finite elements with an edge size of 0.6 mm for forward simulation. The meshes were generated using a MATLAB-based PDE solver. Sixteen electrodes were created around the periphery of the lattice. An open-source MATLAB-based suite, Electrical Impedance and Diffuse Optical Reconstruction Software (EIDORS), was used to compute boundary voltages within the domain. [3]

The solution to the forward problem, using unity bulk conductivity and an adjacent current injection and measurement scheme, is shown in **Figure S1**. Here, the domain potential ϕ across the 2D FE geometry is shown when the unity current is injected at the first pair of electrodes. The forward model is utilised to form the sensitivity or Jacobian matrix (***H***). This matrix relates the changes in the measured voltages at the electrodes to the changes in the conductivities of the finite elements.

The inverse problem is ill-posed and seeks to estimate the conductivity distribution of the domain by minimising the difference in the numerically computed and experimentally measured voltages across the boundary. In this study, we use dynamic imaging, which takes voltage measurement sets at two different time instants and estimates the change in conductivity distribution between the two time instants. [4]

The following equation estimates the conductivity change, given the voltage differences, using the least squares estimate.

$$\mathbf{\Delta\sigma} = (\mathbf{H}^T\mathbf{H})^{-1}\mathbf{H}^T\mathbf{\Delta V} \quad (4)$$

However, this estimate is unstable because of the limited number of independent voltage measurements and $\mathbf{H}^T\mathbf{H}$ being poorly conditione, which amplifies small measurement errors into significant changes in the reconstructed conductivity distribution [4]. Estimations can be improved using regularisation.

The maximum a posteriori (MAP) method determines the solutions as the most likely estimate $\Delta\boldsymbol{\sigma}$ given the measured voltage changes and prior statistical information about the experimental conditions. [4]

$$\Delta\boldsymbol{\sigma} = (\boldsymbol{H}^T\boldsymbol{W}\boldsymbol{H} + \mu\boldsymbol{Q})^{-1}\boldsymbol{H}^T\boldsymbol{W}\Delta\boldsymbol{V} = \boldsymbol{B}\Delta\boldsymbol{V} \tag{5}$$

Here, $\boldsymbol{W}$ is the noise covariance matrix, and is a diagonal matrix. The diagonal elements represent the inverse variance of the voltage measurements. This is quite helpful in case of faulty electrodes. In that case, corresponding diagonal elements can be set to zero, enabling image reconstruction from a reduced data set which does not have the measurements from the faulty electrode(s). In the present study, we assume equal noise in each voltage measurement and thus replace $\boldsymbol{W}$ by the Identity matrix. [5]

The matrix $\boldsymbol{Q}$ is the image covariance matrix, which regularises the estimation of the conductivity distribution. In this study, we employ a Laplace prior, which is controlled by the regularisation hyperparameter $\mu$. The optimal value of the hyperparameter is determined as per the resulting noise figure (NF) of the MAP algorithm, which equals the ratio of the signal-to-noise-ratio (SNR) of the voltage measurements and the SNR of the reconstructed conductivity. We selected the regularisation parameter corresponding to NF equals 1 [4,5].

$\boldsymbol{B}$ is the image reconstruction matrix and can be computed offline. It is evaluated once, and when multiplied by the change in voltage measurement vector, yields the reconstructed conductivity distribution, enabling real-time imaging.

Thus, the reconstruction process requires only a single matrix multiplication and is significantly faster than iterative techniques, facilitating *in situ* image reconstruction. The fractured ligament of the cellular lattice with its corresponding 2D and 3D EIT image reconstruction is shown in **Figure S6.**

**S2. Electrode fabrication and validation of contact stability**

The electrodes were fabricated using a multi-step process designed to minimise variations in contact impedance during mechanical loading. First, the lattice surface was coated with a uniform layer of silver paint (**Figure S3(b)**). A multistrand conductive wire was then wrapped around the ligament, ensuring direct contact with the surface at multiple points, as shown in **Figure S3(c)**. A two-part silver epoxy adhesive (Flexible Silver 16, Epoxy International) was prepared by mixing the epoxy and hardener in a 100:115 weight ratio. This mixture was applied via syringe–needle deposition to the wire and silver-painted region, forming a bead of conductive epoxy on the surface (**Figure S3(d)**). The electrode was then cured at room temperature for 24 hours. This methodology resulted in a mechanically robust and flexible electrical coupling with stable, high electrical conductivity, minimising fluctuations during mechanical loading.

To validate the robustness of the electrode attachment, a rectangular specimen was prepared with one face coated with silver paint. A conductive wire was attached to the silver-painted surface following the procedure described above. A digital multimeter was connected to the specimen, with one probe contacting the silver-painted surface and the other connected to the attached wire. The specimen was secured in the lower grip of a universal testing machine (UTM), while the wire was clamped in the upper grip, as shown in **Figure S4(a)**. The wire was pulled at a displacement rate of 1 mm min$^{-1}$, representing a conservative, worst-case loading condition for the electrode–specimen interface, while the electrical resistance was continuously monitored to detect any changes in contact resistance.

The results show a negligible change in electrical resistance of less than 1 Ω with respect to wire displacement during mechanical loading (**Figure S4(b)**). Notably, failure occurred in the wire between the grips, demonstrating robust electrical contact and strong adhesion between the wire and the specimen.

# FIGURES

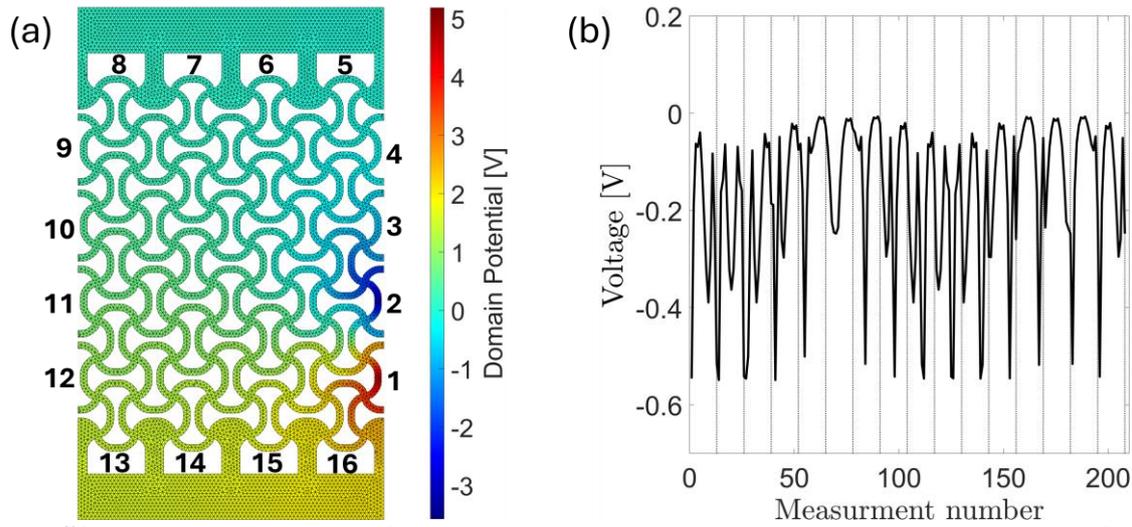

**Figure S2**: (a) Representative forward-problem solution obtained using triangular finite elements, with unity current injected at the first electrode pair. (b) Finite-element-computed voltage differences (208 measurements) for the adjacent current injection and voltage measurement scheme.

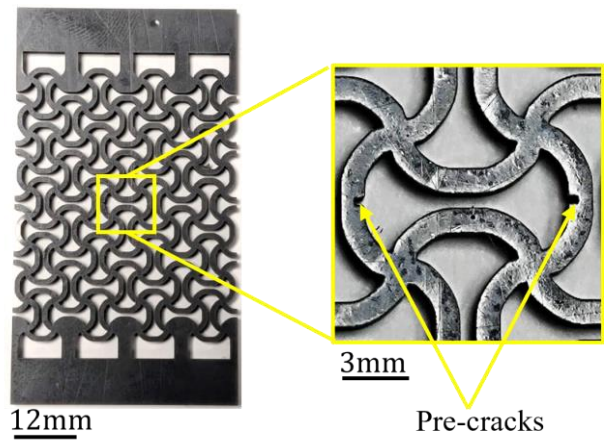

**Figure S2**: Architected lattice with pre-cracks introduced for electrical and mechanical characterisation.

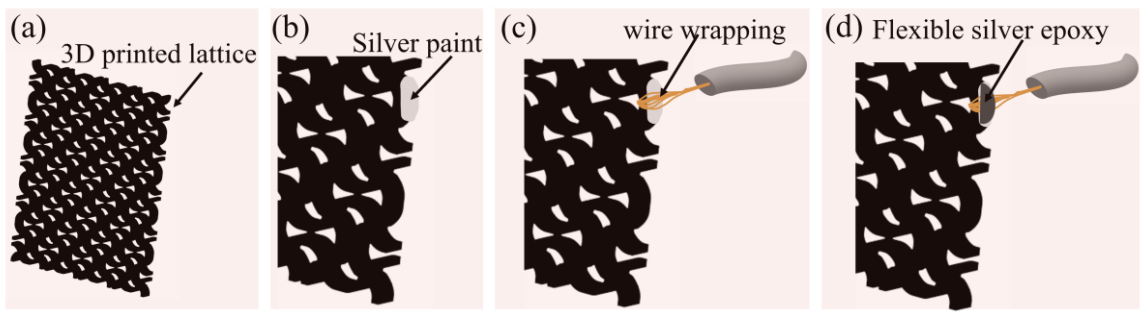

**Figure S3**: Schematic of the multi-step procedure for fabricating electrodes along the periphery of a 3D-printed cellular lattice.

+

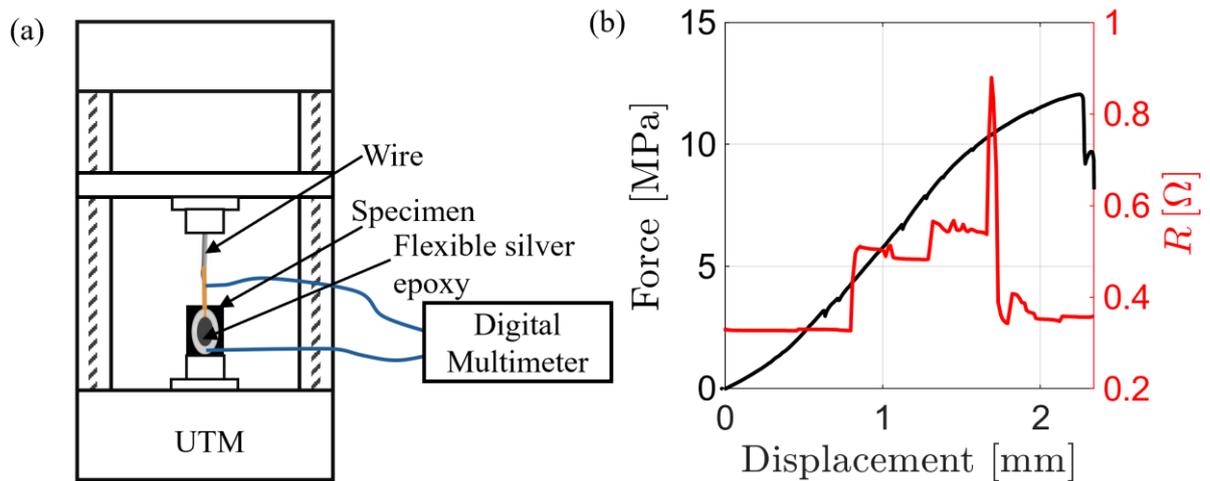

**Figure S4**: (a) Schematic of the experimental setup used to evaluate the robustness of electrode attachment. (b) Applied force and absolute electrical resistance as a function of wire displacement during tensile loading.

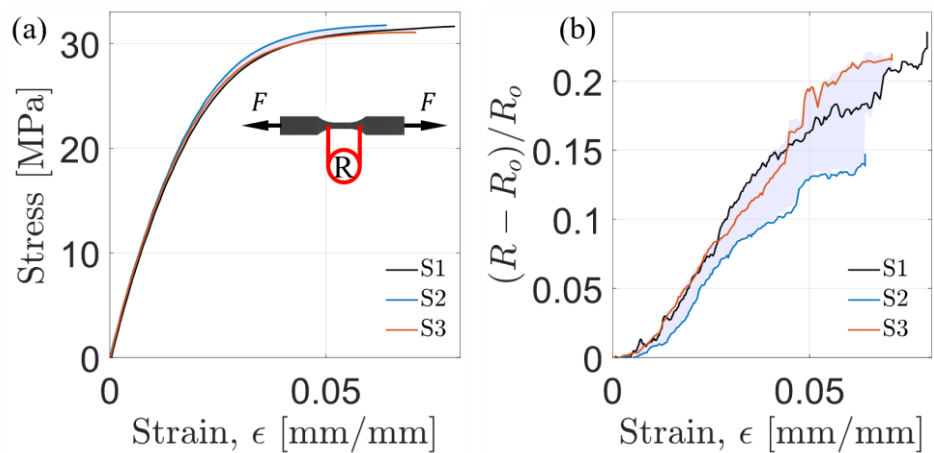

**Figure S5**: (a) Engineering stress–strain response and (b) piezoresistive response of a PG/CNT dogbone specimen under quasi-static tension.

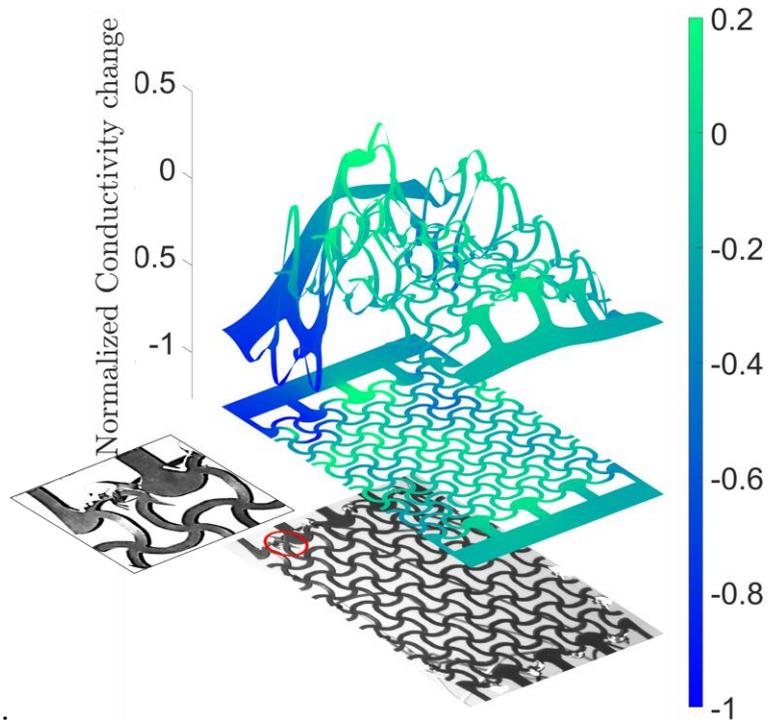

**Figure S6**: Fractured lattice and the corresponding reconstructed EIT image.

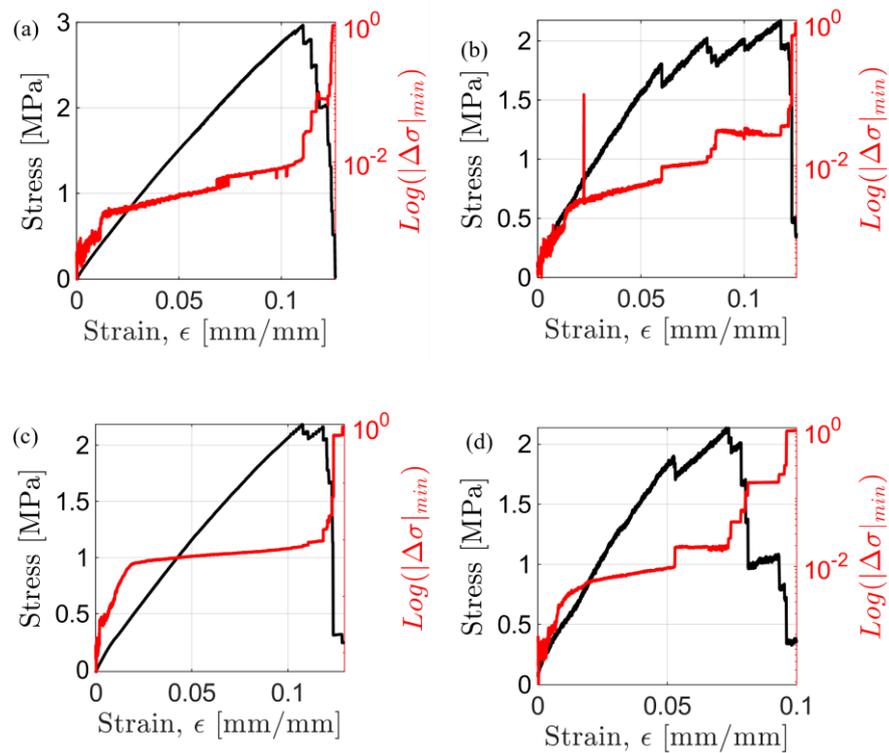

**Figure S7**: Macroscopic stress–strain response and evolution of the logarithmic scaled normalised image amplitude of the lattice for (a) the intact case, (b) the pre-cracked case, (c) simultaneous end-to-end resistance measurement, and (d) EIT with across-current injection.

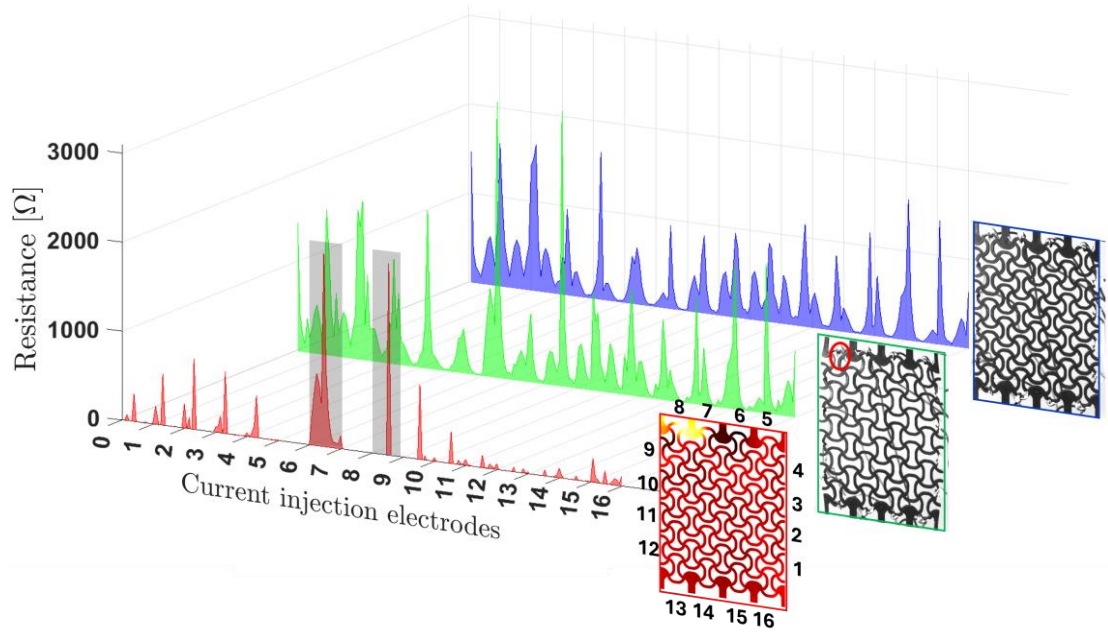

**Figure S8:** Distributions of measured resistances in the unstrained state (violet), after ligament fracture (green), and their difference (red), highlighting localised conductivity loss associated with fracture sites.

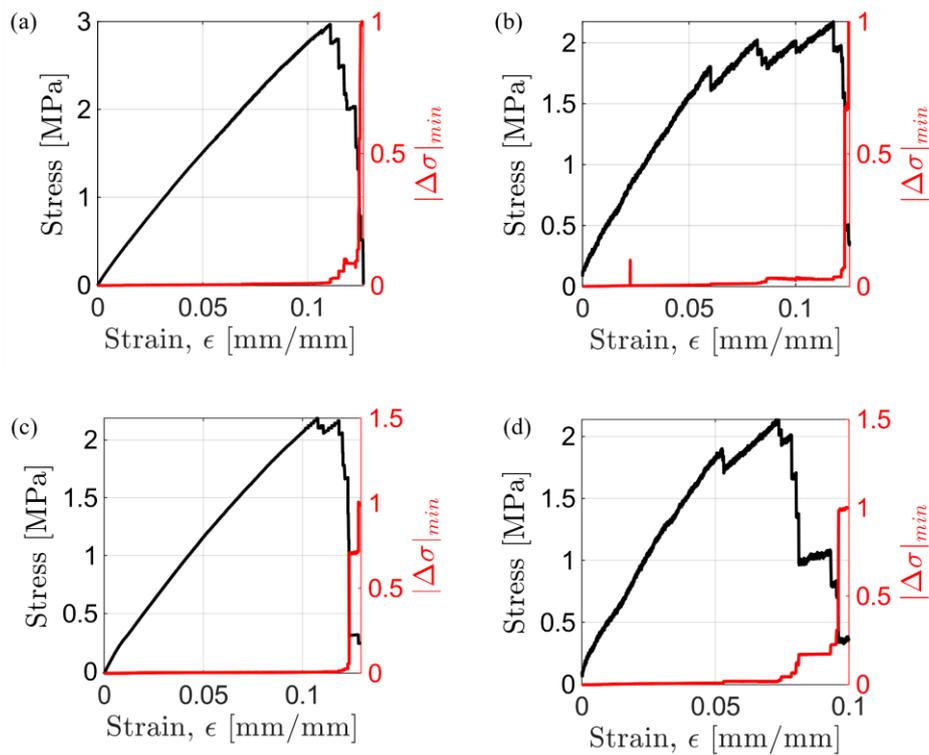

**Figure S9**: Macroscopic stress–strain response and normalised image amplitude evolution of the lattice for (a) the intact case, (b) the pre-cracked case, (c) simultaneous end-to-end resistance measurement, and (d) EIT with across-current injection.

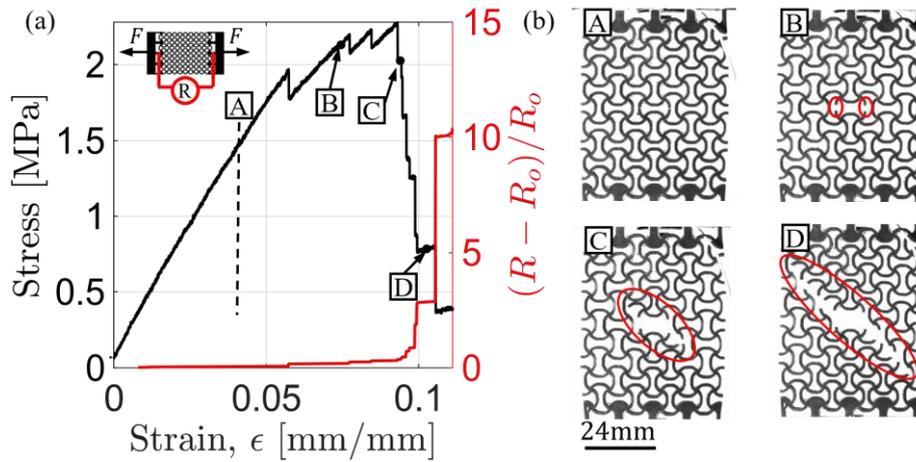

**Figure S9**: (a) Stress–strain and piezoresistive responses of the PG/CNT pre-crack lattice under in-plane quasi-static tensile loading. (b) Deformation and failure maps at different strain levels.

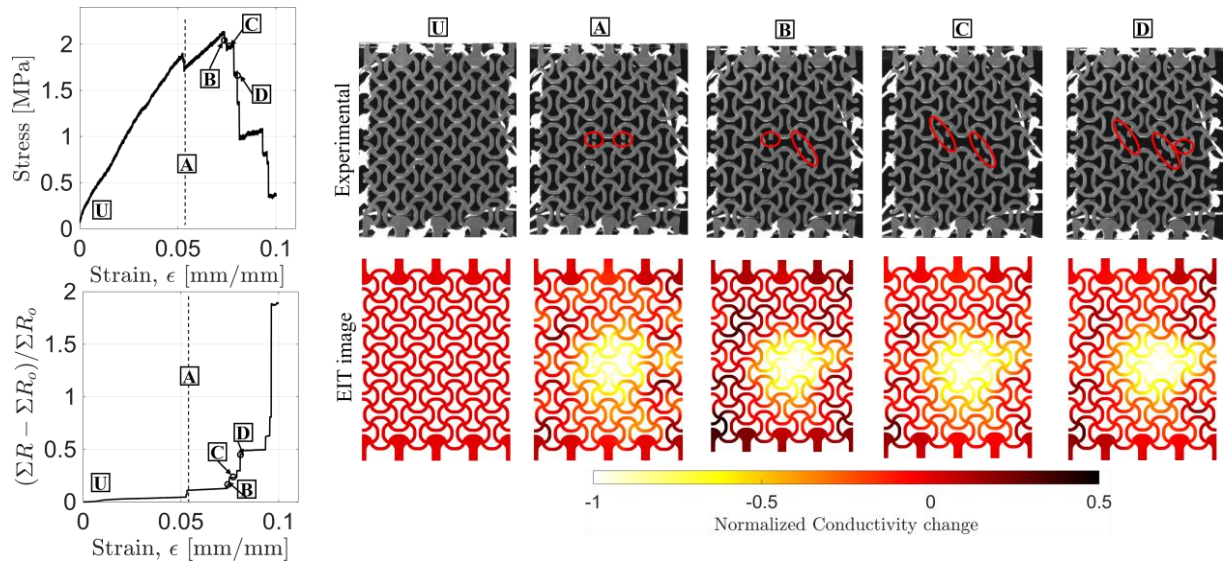

**Figure S10**: Macroscopic stress–strain curve, deformation maps, and EIT maps of the lattice under tension, using opposite-current injection and adjacent voltage measurement schemes.

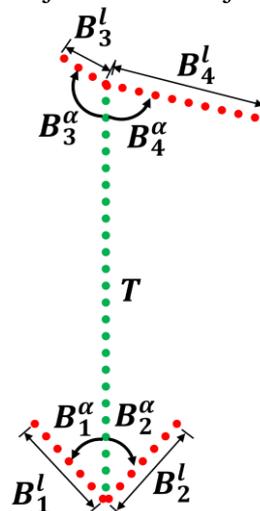

**Figure S11**: Geometric relationship between the branches and the trunk of the motif.

# TABLES

**Table S1**: Design parameters of architected cellular composite geometries explored in this study (See **Figure S11** for parameter definitions)

| Unit cell | Trunk Length (T), mm | Branch 1 | | Branch 2 | | Branch 3 | | Branch 4 | | Unit cell topology |
|---|---|---|---|---|---|---|---|---|---|---|
| | | Length $(B_1^l)$, mm | Angle $(B_1^\alpha)$ | Length $(B_2^l)$, mm | Angle $(B_2^\alpha)$ | Length $(B_3^l)$, mm | Angle $(B_3^\alpha)$ | Length $(B_4^l)$, mm | Angle $(B_4^\alpha)$ | |
| A | 14 | 1.1 | 90 | 1.1 | 90 | 1.1 | 90 | 1.1 | 90 | 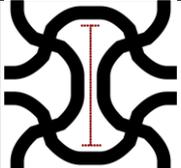 |
| B | 8 | 2 | 50 | 2 | 50 | 5 | 70 | 2 | 110 | 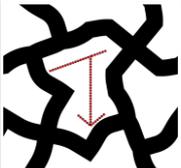 |
| C | 11 | 2 | 70 | 2 | 150 | 2 | 50 | 2 | 130 | 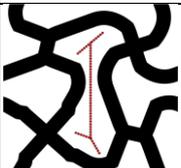 |
| D | 8 | 2 | 110 | 2 | 110 | 2 | 150 | 5 | 70 | 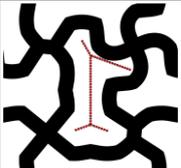 |